\documentclass[prd,groupaddress,nofootinbib]{revtex4}

\usepackage{framed}
\usepackage{bbold}
\usepackage{slashed}
\usepackage{graphicx}
\usepackage{amsmath}
\usepackage{amssymb}
\usepackage{epsfig}
\usepackage{hhline}
\usepackage{soul}
\usepackage{tabularx,pifont,multirow}
\usepackage{enumitem}
\usepackage{colordvi}
\usepackage{soul}
\usepackage{xcolor}
\usepackage{yfonts}

\newcommand{\be}{\begin{equation}}
\newcommand{\ee}{\end{equation}}

\definecolor{darkgreen}{rgb}{0,0.3,0.05}
\makeatletter                                                         %
\newcommand*\rel@kern[1]{\kern#1\dimexpr\macc@kerna}                  %
\newcommand*\widebar[1]{                                              %
  \begingroup                                                         %
  \def\mathaccent##1##2{                                              %
    \rel@kern{0.8}                                                    %
    \overline{\rel@kern{-0.8}\macc@nucleus\rel@kern{0.2}}             %
    \rel@kern{-0.2}                                                   %
  }                                                                   %
  \macc@depth\@ne                                                     %
  \let\math@bgroup\@empty \let\math@egroup\macc@set@skewchar          %
  \mathsurround\z@ \frozen@everymath{\mathgroup\macc@group\relax}     %
  \macc@set@skewchar\relax                                            %
  \let\mathaccentV\macc@nested@a                                      %
  \macc@nested@a\relax111{#1}                                         %
  \endgroup                                                           %
}                                                                     %
\makeatother                                                          %

\begin{document}

\preprint[\leftline{KCL-PH-TH/2018-{\bf 66}}

%

\title{\Large {\bf Spontaneous CPT Violation and Quantum Anomalies in a Model for Matter-Antimatter Asymmetry in the Cosmos
 } }

\bigskip

\author{Nick E. Mavromatos and Sarben Sarkar}

\affiliation{Theoretical Particle Physics and Cosmology Group, Department of Physics, King's College London, Strand, London WC2R 2LS, UK}


\begin{abstract}
\vspace{0.5cm}
\centerline{\bf Abstract }

We review scenarios of baryogenesis through leptogenesis at early epochs of the universe, in string-inspired minimal extensions of the Standard Model (SM), involving heavy right-handed Majorana neutrinos. Spontaneous violation of CPT symmetry is induced by appropriate (in general, temperature-dependent) backgrounds of the Kalb-Ramond (KR) axion field, which has its  origins in the (bosonic) massless string multiplet. As interesting features of the model, we also discuss two issues associated with quantum (chiral) anomalies: (i) the non-contribution of the KR axion background to the (anomalous) chiral magnetic effect, which arises in the presence of external electromagnetic fields and non-zero chiral chemical potentials of charged fermions and (ii) the potential r\^ole of quantum fluctuations of the KR axion on the (anomalous) radiative generation of a Majorana mass for the right-handed neutrinos themselves.

\end{abstract}
\maketitle

\section{Introduction \label{sec:intro}}

It is well known~\cite{Kuzmin:1985mm,Gavela:1993ts,Gavela:1994dt} that the SM of particle physics cannot explain the observed (primarily baryonic)
matter-antimatter asymmetry in the universe~\cite{Planck} which requires:
\begin{equation}
\Delta n(T\sim 1~{\rm GeV})=\frac{n_{B}-n_{\overline{B}}}{n_{B}+n_{\overline{B}}}\sim\frac{n_{B}-n_{\overline{B}}}{s}=(8.4-8.9)\times10^{-11}
\label{basym}
\end{equation}
for (cosmic) times $t\sim10^{-6}$~sec
and temperatures $T\sim1$~GeV. In the above formula $n_{B}$ ($n_{\overline{B}}$)
denotes the (anti-) baryon density in the universe, and $s$ is the
entropy density of the universe, scaling with the cubic power of the temperature. The observation of charge-parity violation led Sakharov to conjecture that fundamental particle interactions~\cite{Sakharov} could lead to Baryon Asymmetry provided there was (i)  Baryon (B) number violation, (ii) Charge (C) and Charge-Parity (CP) symmetry breaking, and (iii) the universe was out of chemical equilibrium (so that the asymmetry between matter and antimatter is not washed out). Sakharov assumed that CPT is a fundamental and unbroken symmetry (where T denotes time-reversal symmetry).
Although the Sakharov conditions are satisfied qualitatively in the SM, 
the calculated baryon asymmetry in the universe (BAU) is found to be \emph{several} orders of magnitude smaller than the observed one (\ref{basym}).

There are two types of non-equilibrium processes
in the early universe that can  produce asymmetries between particles and antiparticles: the first
type concerns processes generating asymmetries between leptons and
antileptons (\emph{leptogenesis})~\cite{Buchmuller:2004nz,Davidson:2008bu,Pilaftsis:2013nqa,Biondini:2017rpb}, while the second produces asymmetries
between baryons and antibaryons directly (\emph{baryogenesis})~\cite{Cohen:1993nk,Trodden:1998ym,Riotto:1999yt,Buchmuller review}.

Several non-trivial extensions of the SM exist, including supersymmetric and (super)string theories and extra dimensional models, which involve extra sources for CP violation so that they could reproduce the observed asymmetry (\ref{basym}), in accordance to Sakharov's conditions. None of the ingredients of those models, though, have been verified by experiment to date. Moreover, fine tuning and some \emph{ad hoc} assumptions are involved in such scenarios, especially in connection with the magnitude of the \emph{CP} violating phases and the associated decay widths. Consequently, the quest for a proper understanding of the observed BAU is incomplete and requires further investigation of physics beyond the SM.

Some of the attempts mentioned above involve the elegant mechanism of baryogenesis via leptogenesis. In such scenarios, a lepton asymmetry is generated first, by means of decays of right handed sterile neutrinos to SM particles, and is subsequently communicated to the baryon sector by means of sphaleron processes which violate both Baryon (B) and Lepton (L) numbers, but preserve the difference B-L~\cite{Fukugita-Yanagida,Luty,Pilaftsis:1997jf,Buchmuller:2005eh,strumia,Shaposhnikov:2006xi}.

Heavy sterile neutrinos, through the the seesaw mechanism~\cite{seesaw}, play another essential r\^ole in particle physics, since they provide a natural explanation for the smallness of the mass of the three light neutrinos in the SM, as suggested by the observed neutrino oscillations~\cite{oscil}. A model which extends the SM in such a way as to understand both neutrino oscillations \emph{and} BAU would be economical and attractive.

 All the above scenarios for the generation of BAU respect the  CPT theorem~\cite{cpttheorem}:  relativistic, local and unitary Lagrangians (without gravity) are invariant under the combined action of C, P and T transformations~\cite{Greenberg,Chaichian}. 
Such a theorem is a cornerstone of modern particle physics and is the reason that it was assumed in the scenario of Sakharov~\cite{Sakharov}. It is still possible that CPT is spontaneously broken.
Also, some of the assumptions in the proof of  the \emph{CPT}  theorem may not necessarily hold in the early universe when quantum gravity effects can be important. If there is CPT violation (CPTV) the necessity of  non-equilibrium processes for the generation of BAU 
in \emph{CPT}  invariant theories, may be relaxed . 
 
In our previous work~\cite{ms,decesare,bms,bms2}  we have considered Lorentz invariance violating (LV) backgrounds in the early universe as a form of \emph{spontaneous} violation of Lorentz  and \emph{CPT}  symmetry.  If LV is the primary source of CPTV, then the latter can be studied within a local effective field theory framework, which is known as the \emph{Standard Model Extension} (SME)~\cite{sme}. From experiments there are very stringent upper bounds on the magnitude of the parameters determining Lorentz and \emph{CPT}  violation in the SME~\cite{smebounds}. 
However, under the extreme conditions of the very early universe, such violations could be significantly stronger than in the present era. We will determine relaxation mechanisms (with temperature), by means of which such strong CPTV at early eras diminishes to phenomenologically acceptable values today.

Within the SME framework, there have been suggestions~\cite{Bertolami} of direct baryogenesis due to LV and CPTV terms in the the effective action, which 
induce ``effective chemical potentials'', say for quarks. In the presence of a chemical potential, the populations of quarks and antiquarks are already different within thermal equilibrium, since the 
the particle  and antiparticle phase-space distribution functions $f(E,\mu), f(\overline E, \overline \mu)$, with $E$ the energy  are different . All these cause a difference in the corresponding equilibrium populations
\begin{eqnarray}~\label{cptvf}
f(E,\mu)=[{\rm exp}(E-\mu)/T)\pm1]^{-1}~, \quad 
f(\overline{E},\bar{\mu})=[{\rm exp}(\bar{E}-\overline{\mu})/T)\pm1]^{-1}~,
\end{eqnarray}
(where the $+ (-)$ will denote a fermionic (bosonic) particle species) and a bar over a quantity associates the quantity with an antiparticle; we have for the chemical potentials
 $\overline \mu = -\mu$. In principle, such scenarios in the SME context, can lead to alternative explanations for the observed matter-antimatter asymmetry, \emph{provided} that detailed mechanisms for freeze-out of particle interactions  are provided. However, in \cite{Bertolami} microscopic models leading to such SME Lagrangians and related phenomena have not been provided.\footnote{If the effects of CPTV manifested themselves only in mass differences between particles and antiparticles, then,
under the natural assumptions~\cite{Dolgov} it has been argued that the dominant effects should come from the quarks; since the quark-antiquark mass differences are bounded from above by the current bounds on the proton-antiproton mass difference~\cite{base}, the  induced BAU, due to the corresponding differences in the thermal distribution functions (\ref{cptvf}), turns out to be several orders of magnitude smaller than the observed value (\ref{basym}). 
This result is reached~\cite{Dolgov} by applying the standard linear scaling of the quark mass with temperature 
in the range of validity of (\ref{basym}).}

In our papers~\cite{ms,decesare,bms,bms2} we have embarked on a study of CPTV induced matter-antimatter asymmetry in the Universe, with a microscopic ultra-violet complete string theory model in mind. Within the framework of effective field theories the string theory model can be mapped into specific SME models. 
In particular, we consider minimal extensions of the SM, involving massive right handed (sterile) Majorana neutrinos (RHN) with a Higgs portal, connecting the RHN sector to the SM one. The RHN couple to the massless Kalb-Ramond (KR) pseudoscalar field (KR axion), dual to the spin-one antisymmetric tensor field of the massless bosonic string multiplet~\cite{gsw}.  The decays of the RHN into SM leptons, in the presence of LV and CPTV backgrounds of the KR axion field, lead to leptogenesis; subsequently  baryogenesis ensues through B-L conserving sphaleron processes in the SM sector. 

 The structure of the article is as follows: in the next section \ref{sec:lepto} we review the basic features of the 
 mechanism for the CPTV-induced lepton asymmetry in a rather generic framework, without specifying the microscopic origin of the CPTV. In section \ref{sec:string}, we discuss microscopic scenarios for such a CPTV-induced matter-antimatter asymmetry, within the framework of effective field theories representing the low-energy limit of strings. In section \ref{sec:cme}, we discuss the r\^ole of the CPTV background on the chiral magnetic effect (CME), 
 which has its origin in the chiral anomaly; CME characterises charged fermion systems such as quarks (effectively massless at sufficiently high temperature) in the presence of external magnetic fields. As we demonstrate, the CPTV KR axion background does not participate in the effect. In section \ref{sec:mass}, we discuss scenarios in which the quantum fluctuations of the KR axion can lead to radiative (anomalous) generation of the RHN mass, which go beyond the standard seesaw mechanisms. Finally, section \ref{sec:concl} contains our conlcusions and outlook for future work.

\section{Spontaneous-CPT-Violation-Induced Leptogenesis \label{sec:lepto}}

In refs.~\cite{ms,decesare,bms} we have discussed the generation of matter-antimatter asymmetry in the universe (in particular baryogenesis via leptogenesis) by invoking spontaneous breaking of CPT symmetry in the early universe through appropriate backgrounds.  
In particular, in \cite{decesare,bms} we have considered leptogenesis originating from tree-level decays of RHN  into SM leptons, in the presence of generic CPTV time-like axial backgrounds. In the cosmological (i.e. Robertson-Walker) frame of the early universe the background was assumed constant. The relevant Lagrangian is given by: 
\begin{equation}
\label{smelag}
\mathcal{L}= {\mathcal L}_{\rm SM} + i\overline{N}\slashed{\partial}N-\frac{m_N}{2}(\overline{N^{c}}N+\overline{N}N^{c})-\overline{N}\slashed{B}\gamma^{5}N-\sum_k \, y_{k}\overline{L}_{k}\tilde{\varphi}N+h.c.
\end{equation}
where ${\mathcal L}_{\rm SM}$ denotes the SM Lagrangian, $B_\mu$ is a CPTV background field, associated with physics beyond the SM, 
$N$ is the RHN field, of (Majorana) mass $m_N$,  $\tilde \varphi$ is the adjoint ($\tilde{\varphi}_i=\varepsilon_{ij}\varphi_j $) of the Higgs field  $\varphi$, 
 and $L_{k}$ is a lepton (doublet) field of the SM sector, with $k$ a generation index. $y_k$ is a Yukawa coupling, which is non-zero and provides a non-trivial (``Higgs portal'') interaction between the RHN and the SM sectors. In the case of \cite{decesare,bms} a single sterile neutrino species 
suffices to generate phenomenologically relevant lepton asymmetry; also, from now on, for simplicity
we restrict ourselves to the first generation ($k=1$) of SM leptons, and set 
\begin{equation}\label{yc1}
y_1 = y ~.
\end{equation}
In the scenario of \cite{decesare,bms}, the CPTV background $B_\mu$ is assumed to have only a non-zero temporal component, which was taken to be constant in the Robertson-Walker frame of the early Universe, 
\begin{equation}\label{temporalB}
B_0 = {\rm const} \ne 0~, \, B_i = 0 ~, i=1,2,3~.
\end{equation}
In this case, the Lagrangian (\ref{smelag}) assumes the form of a SME Lagrangian in a Lorentz and CPTV background~\cite{sme}. 

A lepton asymmetry is then generated due to the CP and CPTV tree-level decays of the RHN  into SM leptons, 
in the presence of the background (\ref{temporalB}), induced by the Higgs portal Yukawa interactions of (\ref{smelag})~\cite{decesare,bms}:
\begin{eqnarray}\label{4channels}
{\rm Channel ~I}&:& \qquad  N \rightarrow l^{-}h^{+}~, ~ \nu \, h^{0}~,  \\ \nonumber 
{\rm Channel ~II}&:& \qquad  N \rightarrow l^{+}h^{-}~,~  \overline \nu \, h^{0}~.
\end{eqnarray}
where $\ell^\pm$ are charged leptons, $\nu$ ($\overline \nu$) are light, ``active'', neutrinos (antineutrinos) in the SM sector,
$h^0$ is the neutral Higgs field, and 
 $h^\pm$ are the charged Higgs fields\footnote{At high temperatures, above the spontaneous electroweak symmetry breaking, the charged Higgs fields $h^\pm$ do not decouple from the physical spectrum, and play an important r\^ole in leptogenesis.}.  As a result of the $B_0 \ne 0$ background (\ref{temporalB}), the decay rates of the RHN between the channels I and II are different, resulting in a lepton asymmetry, $\Delta L^{TOT}$, which then freezes out at a temperature $T_D$. In \cite{bms}, a detailed study of the associated Boltzmann equations for the processes in (\ref{4channels}), and the reciprocal processes,  has led to the result\footnote{There is a certain uncertainty related to the Taylor expansions used in our Pad\'e analysis and to indicate this a range of numbers in brackets is given below.}:
\begin{equation}\label{totDL}
\dfrac{\Delta L^{TOT}}{s} \simeq  (0.016, \, 0.019) \,  \dfrac{B_{0}}{m_{N}}, \quad {\rm at~ freezeout~temperature} \quad T=T_D : \quad m_N/T_D  \simeq (1.44, \, 1.77). 
\end{equation}
 This implies that phenomenologically acceptable values of the lepton asymmetry of ${\mathcal O}(8 \times 10^{-11})$, which can then be communicated to the baryon sector via Baryon-minus-Lepton-number ($B-L$) conserving sphaleron processes in the SM, thus producing the observed amount of baryon asymmetry (baryogenesis)  in the Universe,  occur for values of
\begin{equation}\label{b0}
\frac{B_0}{m_N} \sim  10^{-9}, \quad {\rm at~ freezeout~temperature} \quad T=T_D : \quad m_N/T_D  \simeq (1.77, 1.44),
\end{equation}
The different values $(a,b)$ of the numerical coefficients in the right-hand-side of  the two equations in (\ref{totDL}), are due to two different analytical methods (series expansion ($a$) and integrating factor ($b$) method~\cite{bms}, respectively) used in the Pad\`e approximant solution of the Boltzmann equations associated with (\ref{4channels}). Withe value of the Yukawa coupling (\ref{yc1}) $y \sim 10^{-5}$, and for $m_N = {\mathcal O}(100)$~TeV~\cite{decesare,bms} we thus obtain a $B_0 \sim 0.1~{\rm MeV}$, for phenomenologically relevant leptogenesis to occur at $T_D \simeq (56 - 69) $ TeV, in our scenario.
In \cite{decesare,bms} the microscopic nature of the background $B_0$ was not discussed in detail. 

\section{Microscopic String-Inspired Models \label{sec:string}}

The field strength of the spin-1 antisymmetric tensor  (Kalb-Ramond (KR)) field of the massless (bosonic) gravitational multiplet of strings~\cite{ms,decesare} could provide a simple and physically interesting CPTV background that plays the role of $B_0$ . In the closed-string sector the massless bosonic gravitational multiplet of a string theory consists of three fields~\cite{gsw}: a traceless, symmetric, spin-2  tensor field $g_{\mu\nu}$ identified with the graviton; 
a spin 0 (scalar) field, the dilaton $\Phi$, identified with the trace of the graviton; and the spin-1 antisymmetric tensor (Kalb-Ramond) field  $B_{\mu\nu} = - B_{\nu\mu}$. In the closed string sector, there is a $U(1)$ gauge symmetry 
$B_{\mu\nu} \rightarrow B_{\mu\nu} + \partial_\mu \theta_\nu - \partial_\nu \theta_\mu$ which characterises the target-space effective action. This implies that
the latter depends only on the field strength of the field $B_{\mu\nu}$, which is a three-form with components
\begin{equation}\label{hfield}
H_{\mu\nu\rho} = \partial_{[\mu}\, B_{\nu\rho]},
\end{equation}
where the symbol $[\dots ]$ denotes complete antisymmetrisation of the respective indices. 
The 3-form $H_{\mu\nu\rho}$ satisfies the Bianchi identity 
\begin{equation}\label{bianchi}
\partial_{[\mu}\, H_{\nu\rho\sigma]} = 0, 
\end{equation}
by construction~\footnote{In (Heterotic) string theory, in the presence of gauge and gravitational fields, the right-hand-side of (\ref{hfield}) is modified by appropriate (parity-violating) Chern--Simons three-forms. The right-hand side of the Bianchi identity (\ref{bianchi}) becomes non-zero, a sign of gauge and gravitational anomalies~\cite{gsw}. We shall not deal explicitly with such (higher derivative) terms here, as they are not directly relevant to our leptogenesis scenario. We shall briefly demonstrate, though, in section \ref{sec:cme}, that their inclusion does not affect the chiral magnetic effect.}. 

In the Einstein frame~\cite{string} the bosonic part of the (four-space-time-dimensional) effective action, $S_B$ reads :
\begin{align}\label{sea2}
S_B =&\; \dfrac{1}{2\kappa^{2}}\int d^{4}x\sqrt{-g}\Big(R - e^{-4\Phi}H_{\lambda\mu\nu}H^{\lambda\mu\nu} - \Omega\Big) + \dots,
\end{align}
where  $\kappa^2 = 8\pi G$, and $G= M_P^{-2}$ is the (3+1)-dimensional Newton constant (with $M_P= 1.22 \times 10^{19}$~GeV, the four-dimensional Planck mass). $M_P$ is related to the string mass scale $M_s$ via~\cite{gsw}: ${G}^{-1} = {\mathcal V}^{(n)} \,  M_s^{2+n}$, with ${\mathcal V}^{(n)}$ a compactification volume (or appropriate bulk volume factor, in brane universe scenarios). For standard (ten-space-time dimensional) superstrings $n=6$. 
The $\Omega$ in (\ref{sea2}) represents a vacuum energy term. It arises in non-critical-dimension string models~\cite{aben}, or from bulk contributions in brane universe scenarios; in the latter case, contributions can be of anti-de-Sitter-type~\cite{rizos}. The $\dots$ represent 
derivatives of the dilaton field, $\Phi$; these derivative terms are assumed to be small~\cite{decesare,bms}  
at the epoch of the universe relevant for leptogenesis; hence we may  approximate $\Phi \simeq {\rm constant}$, which can thus be absorbed in appropriate normalisations of the KR field. In this approximation, 
the vacuum energy term $\Omega$ is treated as a constant, to be determined phenomenologically by requiring appropriately suppressed vacuum energy contributions. We shall come back to this issue later. 

It is known~\cite{gsw,string} that the KR field strength terms $H^2$ in (\ref{sea2}) can be absorbed into a generalised curvature scheme with a ``torsionful connection'' ${\overline \Gamma}_{\mu\nu}^{\rho}$ given by ${\overline \Gamma}_{\mu\nu}^{\rho} = \Gamma_{\mu\nu}^\rho +  H_{\mu\nu}^\rho  \ne {\overline \Gamma}_{\nu\mu}^{\rho}$, where $\Gamma_{\mu\nu}^\rho = \Gamma_{\nu\mu}^\rho$ is the torsion-free Christoffel symbol; so the contorsion tensor is determined by the $H_{\mu\nu}^\rho$ field strength.

In our approach we shall include fermion fields, of mass $m$, as well. The contorsion interpretation of $H_{\mu\nu}^\rho$ implies a minimal coupling of this field to the axial fermion current, since the corresponding Dirac term for fermions in torsionful gravitational backgrounds reads~\cite{kaloper,decesare,bms}:
\begin{align}\label{fermions}
S_{Dirac} &= \, \int d^4x \sqrt{-g} \, \Big[ \frac{\imath}{2} \,\Big(\overline \psi \gamma^\mu {\overline {\mathcal D}}(\overline \omega)_\mu \, \psi - ( {\overline {\mathcal D}}(\overline \omega)_\mu \, \overline \psi  )\, \gamma^\mu \, \psi \Big) - m\, \overline \psi \, \psi \Big], \nonumber \\
& =\; \int d^{4}x\sqrt{-g}\bar{\psi}\Big(\imath\gamma^{\mu}\partial_{\mu} - m\Big)\psi + \int d^{4}x\sqrt{-g} \, ({\mathcal F}_\mu + B_{\mu})\, \bar{\psi}\gamma^{5}\gamma^{\mu}\psi~,  \nonumber \\
 {\overline {\mathcal D}}_a  &= \partial_a  - \frac{\imath}{4} \, \overline \omega_{bca}\, \sigma^{bc}, \quad \sigma^{ab} = \frac{\imath}{2}[\gamma^a, \gamma^b]~,
 \nonumber \\
 {\mathcal F}^\mu & =   \varepsilon^{abc\mu} \, e_{b\lambda} \,  \partial_a \, e^\lambda_c ~,  \quad B^\mu = -\dfrac{1}{4}e^{-2\phi}\varepsilon_{abc}^{\;\;\;\;\;\mu}H^{abc}, \quad J^{5 \mu} = \bar{\psi}\gamma^{\mu}\gamma^{5}\psi,
\end{align}
where $e^{a}_\mu (x)$ are the vielbeins, $g_{\mu\nu} (x) = e_\mu^a (x) \, \eta_{ab} \, e_\nu^b (x)$, with $\eta_{ab}$ the Minkowski metric of the tangent space at a space-time point with coordinates $x^\mu$.  The generalised spin-connection is: $\overline \omega_{ab\mu}= \omega_{ab\mu} + K_{ab\mu}$,  $K_{abc} =\frac{1}{2} \, (H_{cab}  - H_{abc} - H_{bca}) = - \frac{1}{2} \, H_{abc}$, where, 
$\omega_{ab\mu}$ denotes the standard torsion-free spin connection; as usual, Latin letters denote tangent-space indices, while Greek letters refer to space-time indices.  In (\ref{fermions}), we used standard 
properties of the $\gamma$-matrices.  For a Robertson-Walker metric $g_{\mu\nu}$ background, of relevance to us here, ${\mathcal F}_\mu =0$, and thus we can write the 
action (\ref{fermions}) in  the form:
\begin{align}\label{fermions2}
S_{Dirac} = &\; \int d^{4}x\sqrt{-g}\bar{\psi}\Big(\imath\gamma^{\mu}\partial_{\mu} - m\Big)\psi + \int d^{4}x\sqrt{-g}\, B_{\mu} \, \bar{\psi} \gamma^{5}\gamma^{\mu}\psi \, \equiv \; S_{Dirac}^{Free} - \int d^{4}x\sqrt{-g}B_{\mu}J^{5\mu}, 
\end{align}
yielding a minimal coupling of the $H_{\mu\nu\rho}$ field to the fermion axial current. 

In four space-time dimensions, the KR three-form $H$ can be expressed in terms of its dual pseudoscalar $b(x)$ (KR ``axion'' ) field~\cite{aben,kaloper}
\begin{align}\label{dual}
\partial^{\mu}b = -\dfrac{1}{4}e^{-2\phi}\varepsilon_{abc}^{\;\;\;\;\;\mu}H^{abc},
\end{align}
where $\varepsilon^{0123} = +1, \; \varepsilon_{0123} = -1$, {\emph etc.} are the elements of the gravitationally covariant (totally antisymmetric) Levi-Civita tensor.
On account of the definition of $B_\mu$ in (\ref{fermions}), this implies 
\begin{equation}\label{bB}
B^{\mu} = \partial^{\mu}b(x)~.
\end{equation}
The total  (four-space-time-dimensional) effective action $S_{eff}$, where we restrict our attention from now on, is given by the sum of the two actions $S_B$ and $S_{Dirac}$,  
\begin{equation}\label{seff}
S_{eff} = S_B + S_{Dirac}
\end{equation}
and  it can be expressed in terms of the KR axion field as follows~\cite{kaloper}:
\begin{itemize} 

\item First, we formulate the path integral, integrated over the KR field strength $H$. 

\item We insist on the preservation of the Bianchi identity (\ref{bianchi}) at a quantum level, via the addition of appropriate counterterms (in a renormalisation group sense) order by order in perturbation theory. This guarantees the conservation of the ``H-torsion charge ''
$Q = \int d^3 x \, \varepsilon_{ijk} H^{ijk}$, which is implemented in the path-integral via a  $\delta$-functional constraint  in the form $\delta\Big(\kappa^{2}\, \varepsilon^{\mu\nu\rho\sigma} \, \partial_{\mu}\, H_{\nu\rho\sigma}\Big), $  
and expressing the latter in terms of a (dimensionless) Lagrange multiplier field $b(x)$, which eventually will correspond to the dual KR axion field: 
\begin{align}\label{delta}
\delta\Big(\kappa^{2}\, \varepsilon^{\mu\nu\rho\sigma} \, \partial_{\mu}\, H_{\nu\rho\sigma}\Big)
&= \int {\mathcal D}b \, \exp\Big[i \, \kappa^{-2}\,\int d^4x \sqrt{-g}\, b(x) \varepsilon_{\mu\nu\rho\sigma} \partial^{\mu } H^{\nu\rho\sigma}\Big] \nonumber \\
&=  \int {\mathcal D}b \, \exp\Big[-i \, \kappa^{-2}\,\int d^4x \sqrt{-g}\, \partial ^\mu b(x) \varepsilon_{\mu\nu\rho\sigma} \,H^{\nu\rho\sigma}\Big]
\end{align}
where the second equality has been obtained by partial integration, upon assuming that the KR field strength dies out at spatial infinity. 

\item Integrating out the $H$-field in the path integral with the action (\ref{seff}), we obtain a path integral over the Lagrange multiplier field $b(x)$, 
\begin{align}\label{seffpi}
Z =&\; \int \, \mathcal{D}g \, \mathcal{D}\psi\, \mathcal{D}\bar{\psi}\, \mathcal{D}b \, \exp[\imath S_{eff}], \nonumber \\
\nonumber
\\
S_{eff} =&\; \dfrac{1}{2\kappa^{2}}\int d^{4}x\sqrt{-g}\,\Big(R + \dfrac{8}{3}\partial_{\sigma} b\, \partial^{\sigma}b - \Omega\Big) 
+ S_{Dirac}^{Free} - \int d^{4}x\sqrt{-g}\partial_{\mu}b\, J^{5\mu}    - \dfrac{3\kappa^{2}}{16}\, \int d^{4}x\sqrt{-g}\,J^{5}_{\mu}J^{5\mu}\Big].
\end{align}

\end{itemize}

In realisitc situations, of relevance to us here, there are many fermion species  $\psi_i$,  with various masses $m_i$, $i = 1, 2, \dots N$. Then the axial current is a sum over such species
\begin{equation}\label{axialcurr}
J_\mu^5 =  \overline \psi_i \gamma_\mu \, \gamma^5 \, \psi_i ~,
\end{equation}
where a repeated index $i$ denotes summation over $i$. 

 In the effective action $S_{eff}$ (\ref{seffpi}), there is a 
four fermion axial-current-current term, which is a \emph{repulsive} four-fermion term, yielding \emph{de-Sitter type} (positive) contributions to the vacuum energy. Such positive contributions are 
standard in Einstein-Cartan theories of quantum torsion, where the torsion can be integrated exactly in a path integral.

On considering KR-axion backgrounds $\bar b(x)$ linear in cosmic time $t$, $\bar b \propto t$, for which ${\dot {\bar b}}  \equiv B_0$ is 
constant (in the Robertson-Walker frame), and which are known to constitute exact  backgrounds in bosonic non-critical strings~\cite{aben}, 
 we observe that the $\partial \bar b$-$J^5$ interaction term in (\ref{seffpi}) yields the CPT-Violating axial background $B_0$-term of the model discussed in \cite{decesare,bms}, which leads to leptogenesis. In this way one obtains a microscopic origin of $B_0$ in the context of string-inspired models. We mention at this point that, as it turns out~\cite{decesare,bms}, the contributions of $B_0^2$ to the vacuum energy density, arising from the kinetic terms of the KR axion field, are too large during the leptogenesis era (since the $B_0 $ required for phenomenologically acceptable lepton asymmetry
is in the range of 0.1 MeV (\ref{b0})). In order for the standard cosmology not be affected significantly, one must require a fine tuning between $B_0^2$ and the bulk-induced vacuum energy density $\Omega$ (which receives anti de Sitter contributions from the bulk~\cite{decesare,bms}), so that the total vacuum energy density acquires acceptably small values during the leptogenesis and subsequent eras. 

A caveat to the above ideas is that $b$-axion backgrounds  linear in time  may not be exact solutions for superstrings in the presence of fermions. Moreover, even if they are,  it is not known whether one could fine tune the associated parameters so as to guarantee a $B_0$ background (\ref{bB}) in the MeV or lower range, as required for leptogenesis  in the scenario of \cite{decesare}, given that a natural mass scale for such backgrounds is provided by the string scale itself $M_s \gg $~MeV~\cite{aben}. 

In \cite{decesare,bms}, an additional case for obtaining a CPTV KR axion background,  corresponding to a constant $B_0 = {\dot {\bar b}}$,
which could lead to low $B_0$ appropriate for leptogenesis,
was proposed, involving fermionic axial condensates, that have been conjectured to occur at the freezeout epoch of leptogenesis. 
Indeed, in the presence of fermions, the equations of motion for the KR background field $\bar b$ from (\ref{seffpi}), imply:
\begin{align}\label{beom}
\partial_{\alpha}\Big[\sqrt{-g}\Big(\dfrac{8}{3\kappa^{2}}\partial^{\alpha}\bar{b} - J^{5 \; \alpha}\Big)\Big] = 0.
\end{align} 
On assuming a (constant) temporal chiral condensate (which respects isotropy of the Universe),
\begin{align}\label{condens}
0 \ne {\rm const}. = \langle J^{05} \rangle = \langle  \psi^\dagger_i \, \gamma^5 \psi_i \rangle~, 
\end{align}
which may characterise fermions in the model except Majorana neutrinos~\cite{decesare}, e.g. quarks in the SM sector, expanding the current 
in (\ref{beom}) about the condensate (\ref{condens}), $J^{5}_{0} = \langle J^{5}_{0}\rangle + $ quantum fluctuations, and ignoring the fluctuations, 
we then obtain from (\ref{beom})
\begin{align}\label{tempB0}
\partial_{t}\Big[\sqrt{-g}\Big(\dfrac{8}{3\kappa^{2}}B^{0} - \langle J^{0\, 5} \rangle \Big)\Big] = 0,
\end{align}
which admits as a consistent solution (\emph{cf.} (\ref{bB}))
\begin{align}\label{b0cond}
B^0 = {\dot {\bar{b}}} = \frac{3\kappa^{2}}{8}\, \langle J_{0\, 5} \rangle = {\rm const.} \ne 0, 
\end{align} 
implying a 
constant (in the Robertson-Wallker frame) Lorentz- and CPT- Violating  axial background $B^0$, as required for leptogenesis in the scenario of \cite{decesare,bms}. 

In view of the LV and CPTV nature of $B_0$, it must satisfy the current-era stringent upper bounds, imposed by a plethora of precision measurements~\cite{smebounds}, according to which $|B_0| < 0.01$~eV (with much more stringent constraints for spatial components $|B_i | < 10^{-31}$~GeV). 
In the constant $B_0$ scenario of \cite{decesare}, this could be guaranteed, if one assumes that the chiral current condensate $\langle J^{05}\rangle$, is destroyed (due to unknown (beyond the SM) physics) at a temperature near the lepton-asymmetry freezeout $T \simeq T_D \simeq 10^5$~GeV.
In that case, from (\ref{tempB0}), upon taking into account a Robertson-Walker space-time, with scale factor $a(t) \sim T^{-1}$ at high temperatures, we obtain a cooling `law' for $B_0 \sim T^3$, for $T \le T_D$, which comfortably satisfies the above constraints in the current epoch~\cite{decesare}, where the average temperature of the universe is that of the Cosmic Microwave Background (CMB) radiation, $T_0 \sim T_{\rm CMB} \simeq 0.23$~meV. Indeed, with such a cooling law, taking into account that at decoupling $B_0(T_D \simeq 10^5~{\rm GeV}) ={\mathcal O}(0.1~{\rm MeV})$, one finds~\cite{decesare}: $B_0 (T_0) = {\mathcal O}(10^{-57})$~GeV, which satisfies comfortably the bounds in \cite{smebounds}. Moreover, the corresponding vacuum energy density contributions, of order $B_0^2$, also satisfy the cosmological constraints~\cite{planck}. 

Although appealing, the above scenario suffered from the fact that no concrete model was proposed for the formation of the axial condensates. Since the four fermion axial interaction terms in (\ref{seffpi}) are \emph {repulsive}, they do not lead to condensate formation. Hence one can invoke other mechanisms beyond the SM, \emph{e.g.} through the appropriate exchange of heavy states that may exist in string theory models. However, such mechanisms have not been elaborated further in \cite{bms}. We are also currently agnostic as to the microscopic mechanism leading to the disappearance of the condensate soon after the freezeout; the $B_0$ background drops with the (cubic power of the) temperature, as time evolves~\cite{decesare,bms}, in order for the model to be compatible with the current stringent bounds on CPTV~\cite{sme}. 

However, in \cite{bms2} we demonstrated that one does not actually need the formation of axial condensates in order to obtain phenomenologically acceptable leptogenesis. In that work, by actually considering \emph{non constant}, temperature-dependent backgrounds $B_0$, obtained from the antisymmetric tensor field of string theory, scaling with the temperature as $T^3$,  as described above, one can still produce a lepton asymmetry during the leptogenesis  era via the decays of the right-handed neutrinos, with a phenomenology similar to the one in constant $B_0$ backgrounds~\cite{decesare,bms}. 
An analytical approximation to the solution of the Boltzmann equations for that case, extending the study in \cite{bms}, has been considered in \cite{bms2},
. 
The cubic dependence on temperature of the background $B_{0}(T) \sim T^{3}$, is dictated by the equation of motion of the KR-axion field (\ref{beom}) \emph{in absence} of an axial-fermion-current condensate, and is 
assumed all the way from temperatures around decoupling till the present day. Such a scaling 
is sufficiently mild, for the high temperature regime that we have considered, so that the conditions for leptogenesis considered in \cite{bms} are only slightly modified. Leptogenesis still occurs at decoupling temperatures of order $T_D \simeq 100$ TeV, where the background field $B_0 = {\mathcal O}({\rm keV})$ is smaller than the one predicted in \cite{decesare,bms}. Nonetheless, we obtain for the current-epoch the range of values $B_0 (T_0) < 3 \times 10^{-57}$~GeV. The upper bound is of the same order as found in the scenario of \cite{decesare}, and lies comfortably within the stringent current bounds of CPTV and LV~\cite{smebounds} as well as the cosmological constraints on the vacuum energy density~\cite{planck}.

\section{KR axion Backgrounds, Anomalies and the Chiral Magnetic Effect \label{sec:cme}} 

In the presence of external magnetic fields, we will briefly discuss potential effects of the axial torsion background $B_0$ on physical phenomena. Primordial magnetic fields are known to play a r\^ole in leptogenesis scenarios~\cite{maglept}. Specifically, 
we would be interested in examining whether $B_0$ plays any r\^ole in the so-called Chiral Magnetic Effect  (CME)~\cite{cme}. The CME has been conjectured to characterise systems (such as neutron stars or a hot QCD quark-gluon plasma ~(QG)) with external magnetic fields in the presence of a chiral chemical potential $\mu_5$\footnote{A chiral chemical potential has different values of the chemical potential for left and right chiral spinors.}.  

As we shall see, though, in our case, the CME will be \emph{unaffected} by the presence of the axial KR background $B_0$, which in this respect plays a r\^ole analogous to an external axial vector potential that is known not to contribute to CME~\cite{kaplan,dvorn}. The non-contribution of the $B_0$ field to the CME in our case should also be expected from: (i) the fact that the phenomenon
 has its origin~\cite{cme} in the chiral anomalies in quantum field theories~\cite{adler},  (ii) the r\^ole 
of the $B_0$ field as a  torsion in the low-energy string effective action, and (iii) the well-known result~\cite{hull} that torsion contributions to the anomaly equation are removable by the addition of appropriate local counterterms (in a renormalisation group sense) to the corresponding effective action. Physical effects, such as the CME, should thus be free from such ambiguities. This result can also be understood from the fact that the chiral anomaly is associated with the index of the Dirac operator for fermions, which the torsion does not contribute to (as can be shown explicitly using heat kernel or other techniques~\cite{mavindex}). 

\subsection{Chiral Anomalies and the Chiral Magnetic Effect}

It will be instructive to first review briefly the CME phenomenon in the QG case~\cite{cme}. Consider the (3+1)-dimensional flat space-time, finite-density \emph{massless} quark Lagrangian ${\mathcal L}_{\rm quartks}$ in the presence of a finite chemical potential $\mu$ and a finite chiral chemical potential $\mu_5$ 
\begin{align}\label{chirallag}
{\mathcal L}_{\rm quartks} \,  \ni \, \int \, d^4 x \, \Big(\mu \sum_{i={\rm quarks}} \, {q}_i^\dagger \, q_i  + \mu_5 \sum_{i={\rm quarks}} \, 
{q}_i^\dagger \, \gamma^5 \, q_i  \Big).
\end{align}
The chiral anomaly implies that the the corresponding chiral current density $J^{5\, \mu}$ is \emph{not conserved}, but its divergence is given by the so-called axial~anomaly~\cite{adler}, which in the case of interest is restricted only to include electromagnetic terms with Maxwell field strength $F_{\mu\nu}= \partial_\mu\, A_\nu - \partial_\nu \, A_\mu$ (with  $A_\mu$ denoting the U(1) gauge potential, corresponding to the photon field), and its dual $\widetilde F_{\mu\nu} = \frac{1}{2}\, \epsilon_{\mu\nu\alpha\beta}\, F^{\alpha\beta}$ (with $\epsilon^{0123}=+1$ in our conventions): 
\begin{align}\label{anomaly}
\partial_\mu J^{5\, \mu} = \frac{e^2}{8\, \pi^2} \, F_{\mu\nu} \, \widetilde F^{\mu\nu} = \frac{e^2}{2\, \pi^2} \, \vec E \cdot \vec {\mathcal B}~,
\end{align}
where $e$ is the electron charge, and $\vec E$ ($\vec {\mathcal B}$) is the electric (magnetic) field respectively, which will be taken to be external in our  discussion. 

Integrating over 3-space, we may rewrite (\ref{anomaly}) in terms of the rate of change of the \emph{chirality} $N_5 = N_R - N_L$~\cite{cme}:
\begin{align}\label{chiral2}
\mu_5 \, \frac{d N_5}{dt} = \frac{e^2\, \mu_5}{2\, \pi^2} \, \int d^3x \, \vec E \cdot \vec {\mathcal B}~.
\end{align}
In arriving at (\ref{chiral2}) we took into account  that the chiral chemical potential $\mu_5$ is the energy required to change a left handed fermion into a right handed one, which equivalently is the energy required to move a particle from the left handed Fermi surface and place it onto the right-handed one. If $\mu_{L(R)} = \mu \mp \mu_5$ denotes the corresponding chemical potentials of the left(right) handed fermions, the above process  costs an energy~\cite{cme} $\mu_R - \mu_L= 2\mu_5$, and this will change the chirality $N_5$ by 2. For an infinitesimal change $dN_5$ of the chirality then, the corresponding cost in energy is given by $\mu_5\, dN_5$, whose rate is then given by (\ref{chiral2}). 
Conservation of energy, implies that this amount must be compensated by the power of the electric field present in the system, which in terms of the electric current density $\vec j_E$  is provided by $\int d^3x \, \vec j_E \cdot \vec E$, thereby leading (on account of (\ref{chiral2})) to:
\begin{align}\label{chiral3a}
\int d^3x \, \vec j_E \cdot \vec E = \mu_5 \, \frac{d N_5}{dt} = \frac{e^2\, \mu_5}{2\, \pi^2} \, \int d^3x \, \vec E \cdot \vec {\mathcal B}~,
\end{align}
from which the CME follows~\cite{cme}, namely the existence of an electrical current proportional to the magnetic field strength and $\mu_5$:
\begin{align}\label{chiral3}
\vec J_E =  \frac{e^2\, \mu_5}{2\, \pi^2} \, \int d^3x \, \vec {\mathcal B} = \frac{2\, \alpha}{\pi} \, \mu_5\, \int d^3x \, \vec {\mathcal B}~,
\end{align}
where the appearance of the fine structure constant $\alpha = e^2/4\pi$ as a proportionality factor indicates the quantum nature of the phenomenon, consistent with its origin from the chiral anomaly~\cite{adler}. In addition, there is an induced chiral current, which is proportional to the chemical potential $\mu$~\cite{cme,mz}:
\begin{align}\label{cme5}
{\vec J}^{\,5} = \frac{e\, \mu}{2\, \pi^2} \, \int d^3x \, \vec {\mathcal B}~.
\end{align}
These effects has been defined in several independent ways in \cite{cme}, including finite temperature $T \ne 0$ formulations, and thus it was argued that CME is independent of temperature for $T$-independent $\mu_5$ and $\mu$. 
Such an effect might have important phenomenological implications for the QG  physics~\cite{cme}. 

\subsection{Non-contribution of the KR Background to the Chiral Magnetic Effect}

In string effective actions~\cite{string,kaloper}, the interpretation of the (totally antisymmetric) $H$-field  as torsion  is valid in an expansion in powers of the Regge slope $\alpha^\prime$ to first order (quartic in derivatives). 
For our cosmological case a torsionful curved space-time is relevant; in such a space-time any chiral anomaly that might characterise our system will also involve the generalised Riemann curvature tensor ${\overline R}_{\mu\nu\rho\sigma}(\overline \omega)$ and its dual~\cite{kaloper}~\footnote{We note that this is the complete form of the anomaly in our case, which is characterised by the
conservation of the H-torsion charge (\ref{delta}), at a quantum level. Indeed, as discussed in \cite{zanelli}, for a generic torsion three form $\mathbf{T} 
= \mathbf{d} \mathbf{e}^a + \overline \omega^a_b \wedge \mathbf{e}^b = \mathbf{K}^a_b \wedge \mathbf{e}^b$, with $\mathbf{e}^a$ the vielbein one-form and  $\mathbf{K}$ the contorsion tensor, the right hand side of the anomaly (\ref{anomalygrav}) also contains the Nieh-Yan topological invariant density~\cite{nie}, 
${\mathcal N} =  {\mathcal N} = \mathbf{T}^a \wedge \mathbf{T}_a -  \mathbf{\overline R} (\overline \omega)_{ab}\, \wedge  \mathbf{e}^a \wedge \mathbf{e}^b  = \mathbf{d} (\mathbf{e}^a \wedge \mathbf{T}_a)$.  In our case, $\star \mathbf{H} \propto \mathbf{e}_a \wedge \mathbf{T}^a $, and thus 
the Nieh-Yan invariant ${\mathcal N}$ vanishes identically on account of the torsion charge conservation constraint (\ref{delta}), which can be written as 
$ 0 = \mathbf{d} \star \mathbf{H} = \mathbf{d}(\mathbf{e}_a \wedge \mathbf{T}^a)= {\mathcal N}$.}
\begin{align}\label{anomalygrav}
\nabla_\mu J^{5\, \mu} = \frac{e^2}{8\, \pi^2} \, F_{\mu\nu} \, \widetilde F^{\mu\nu} - \frac{1}{192\, \pi^2} {\overline R}_{\mu\nu\alpha\beta}(\overline \omega)\, {\widetilde {\overline R}}^{\mu\nu\alpha\beta} (\overline \omega) \equiv {\mathcal G}(A, {\overline \omega})~,
\end{align}
where the overline over a quantity denotes the presence of torsion, and $\overline \omega = \omega + H$ denotes (schematically) the torsionful connection, with $\omega$ the torsion-free connection and $H$ the KR field strength which plays the r\^ole of (totally antisymmetric) torsion. The quantity 
$\nabla_\mu$ denotes the gravitational covariant derivative, with respect to the \emph{torsion-free} connection\footnote{The reader should notice that there is no $H$-torsion contribution to the covariant four-divergence of a four-vector.} .
Tensor duals are defined as $\widetilde F_{\mu\nu} = \frac{1}{2} \, \sqrt{-g}\, \varepsilon_{\mu\nu\rho\sigma} \, F^{\rho\sigma}$,  and 
${\widetilde {\overline R}}_{\alpha\beta\mu\nu} = \frac{1}{2} \, \sqrt{-g} \, \varepsilon_{\mu\nu\rho\sigma} \, {\overline R}_{\alpha\beta}^{\,\,\,\,\,\,\,\,\rho\sigma}$,  
where $g$ is the determinant of the torsion-free metric corresponding to a (torsion-free) Riemann curvature tensor $R_{\mu\nu\rho\sigma}$.

 The gravitational part of the anomaly (in differential form notation) is given by 
\begin{align}\label{rrtorsion} {\rm Tr}\, \Big(\mathbf{\overline R} (\overline \omega) \wedge \mathbf{\overline R} (\overline \omega) \Big)= 
{\rm Tr}\, \Big(\mathbf{R }(\omega) \wedge \mathbf{R }(\omega)\Big) + {\rm \mathbf{d}}\, \Big[{\rm Tr} \Big(\mathbf{H} \wedge \mathbf{R}) + \mathbf{H} \wedge \mathbf{D} \, \mathbf{H} + \frac{2}{3} \, \mathbf{H} \wedge \mathbf{H} \wedge \mathbf{H} \Big)\Big], 
\end{align}
where $\mathbf{d}$ is the exterior derivative, $\mathbf{D}$ denotes the (torsion-free) gravitational covariant exterior form, 
$\mathbf{D}\, \mathbf{V}^a = \mathbf{d} \mathbf{V}^a + {\omega}^a_{\,\,b}\, \wedge \, \mathbf{V}^b$, 
and the trace  Tr is taken over tangent space (Latin) indices $a,b, ...$, i.e. ${\rm Tr}\, \Big(\mathbf{\overline R} (\overline \omega) \wedge \mathbf{\overline R} (\overline \omega) \Big)= \, \mathbf{{\overline R}}^a_{\,\,b}(\overline \omega) \wedge \mathbf{{\overline R}}^{b}_{\,\, a}(\overline \omega)$, with $\mathbf{{\overline R}}^{a}_{\,\,b} (\overline \omega) = \frac{1}{2} {\overline R}_{\mu\nu\,\,\,\,b}^{\,\,\,\,\,\,a}\, dx^\mu \wedge dx^\nu = \mathbf{d}\, {\overline \omega}^a_{\, b} + {\overline \omega}^a_c\, \wedge {\overline \omega}^c_{\, b}$,  etc.,   where the indices $a,b,c \dots$ are raised and lowered by the Minkowski metric.
On the other hand, in order to maintain the conventional $U(1)$ gauge invariance, the Maxwell field strength is defined as in standard electrodynamics~\cite{kaloper}, with respect to the usual derivative ( i.e.
$\mathbf{F}= \mathbf{d A}$, obeying the Bianchi identity 
$\mathbf{d F} =0$).

It is well known~\cite{hull} that the torsion contributions (\ref{rrtorsion}) to the anomaly (\ref{anomalygrav}) 
may be removed by the addition of local renormalisation counterterms to the effective action, provided that the chiral current couples to a gauge field, which is the case of interest here. The anomaly becomes dependent only on the torsion-free spin connection $\omega$~\footnote{As already mentioned, this can also be understood from the fact~\cite{hull,mavindex} that the chiral anomaly is associated with the index of the Dirac operator for fermions, and the latter is not affected by torsion, as it is associated with the topological quantity $\int_{{\mathcal M}} {\rm Tr}\Big(R(\omega) \wedge R(\omega)\Big)= \int_{{\mathcal M}}  {\rm Tr}\Big( {\overline R}(\overline \omega) \wedge {\overline R}(\overline \omega)\Big)$, where ${\mathcal M}$ is a compact target-space-time manifold without boundary, and we took into account that the torsion contributions to the gravitational part of the anomaly (\ref{rrtorsion}) is a total differential.  The latter property would also imply that one can appropriately redefine the axial current by such torsion dependent terms~\cite{dobado}, to arrive at a new gauge and Lorentz-invariant axial current, whose anomaly equation is torsion free.}, 
\begin{align}\label{anomalygrav2}
\nabla_\mu J^{5\, \mu} = \frac{e^2}{8\, \pi^2} \, F_{\mu\nu} \, \widetilde F^{\mu\nu} - \frac{1}{192\, \pi^2} {R}_{\mu\nu\alpha\beta}(\omega)\, {\widetilde {R}}^{\mu\nu\alpha\beta} (\omega) \equiv {\mathcal G}(A, {\omega})~.
\end{align}

This implies a specific form of the low-energy effective action, which we restrict our attention to in this work. 
Indeed, as shown in \cite{kaloper}, and discussed briefly in section \ref{sec:intro}, imposing 
the constraint on the conservation of the torsion charge (\ref{delta}),  is equivalent to the addition of specific counterterms, which leads to the dual effective action (\ref{seffpi}) in terms of the (Lagrange multiplier) KR-axion field $b(x)$. Thus, a QED  effective action (in the concrete case the fermions are charged under electromagnetism)  in a space-time with a torsionful connection, is equivalent to a QED action (\ref{seffpi}) in a space time without torsion but with a dynamical KR axion field. The latter contains a dimension six four-fermion operator and a dimension five operator that couples the derivative of the $b$-field to the axial fermion current. By partial integration, then, this dimension-five term in the effective lagrangian yields a coupling of the $b$ field to the anomaly ${\mathcal G}(A, {\omega})$ (\ref{anomalygrav2}) 
without torsion. Since the torsion contributions to the anomaly are 
removable by an appropriate choice of counterterms, one should not expect any contribution of $B_0$ to the CME, which as discussed above (\emph{cf.} (\ref{chiral2}), (\ref{chiral3})), is linked to the chiral anomaly. Indeed $B_0$ as an axial background, is known not to contribute to CME~\cite{kaplan,dvorn}, as we now review 
briefly, for completeness.

To this end, we first remark that, for a Robertson-Walker (RW) cosmological backgound, the $R \widetilde R$ term in (\ref{anomalygrav2}) \emph{vanishes} identically~\footnote{In fact, it is only the gravitational-wave type fluctuations that contribute to the (torsion-free) Riemann-curvature-dependent part of the anomaly (\ref{anomalygrav}), which can then lead to interesting scenarios for leptogenesis, different from our approach here~\cite{stephon}. Moreover, graviton fluctuations in the $R(\omega) \, \widetilde R(\omega) $ gravitational parts of the anomaly (\ref{anomalygrav}) might play an important r\^ole in radiative Majorana mass generation for the right-handed neutrinos~\cite{pilaftsis}, as we shall review in section \ref{sec:mass}.}. Hence, for cosmological RW space-times the axial anomaly is determined only by its gauge-field part. One can therefore discuss the CME in our context by a straightforward extension of the flat space-time case. 

It suffices for our purposes, to restrict our attention to a local frame, where the expansion of the Universe can be ignored. (This would be the case if one is interested in examining the effects of $B_0$ on CME during the leptogenesis era, or in a quark-gluon (QG) plasma~\cite{cme,kaplan} or in a neutron star~\cite{dvorn2}). In the absence of an explicit $\mu_5$ term,  
we observe from the effective action (\ref{seffpi}), that, at least formally, the background field $B_0 = {\dot {\bar b}}$ seems to play a r\^ole analogous to a (generally temperature dependent) chiral chemical potential. 
If one adds a chemical potential $\mu_5$ term (e.g. in order to capture effects local in space-time in a QG plasma~\cite{cme}), this will appear in the effective action for fermions as the combination 
\begin{align}
\mu_5^{\rm eff} \equiv \mu_5 - B_0(T),
\label{effcp}
\end{align}
which has the apparent form of an {\emph effective} chemical potential. One would thus \emph{naively} expect a CME (\ref{chiral3}), with $\mu_5$ replaced by $\mu_5^{\rm eff}$ (\ref{effcp}). 

However, as argued in \cite{kaplan,dvorn}, using different methods, the axial vector potential $B_0$ does \emph{not} contribute to CME, and instead one has (\ref{chiral3}), even if $B_0 \ne 0$ is present
~\footnote{We should stress, that both the present and previous works of ours~\cite{bms2} pertain to cases in which the axial background is an independent field, e.g. the KR axion. We do not discuss here situations where the totally antisymmetric torsion is a chiral \emph{condensate} of fermions. In such cases, as we remarked in \cite{bms2}, the free-fermion analysis  of \cite{dvorn} needs to be modified to take proper account of the fermion self-interactions in \eqref{seffpi}. For a discussion along those lines, and the potential r\^ole (in the early universe) of torsion arising from thermal condensates of massless chiral fermions, the reader is referred to ref.~\cite{dolan}.}. The subtlety lies in the fact that, in the presence of a background field $B_0$, as we have discussed in \cite{decesare,bms,bms2}, the dispersion relations for the fermions are affected non-trivially by the presence of $B_0$, which differentiates it from the chiral chemical potential case; moreover, there are subtleties related to the order of taking the massless limit $m \to 0$. 
In the presence of a chiral chemical potential, an external constant magnetic field and a (generic, but constant) axial background (of which our (constant) $B_0$ is a special case) the CME was discussed in \cite{dvorn}, within the framework of relativistic quantum mechanics~\cite{rqm}.  
It turns out to be important to take the massless (chiral fermion) limit $m \to 0$ at the end of the computation: one should assume massive fermions, in the presence of a $B_0 \ne 0$, solve the corresponding Dirac equation, and only at the end take the limit $m \to 0$. Had one started, instead, with the chiral Lagrangian for {\emph massless} quarks 
and then turned on an external vector time-like potential, $B_0 \ne 0$, the appearance of $B_0$ contributions to the CME (through (\ref{effcp}))  would have occurred, which, however, would not be correct  from the point of view of energy conservation~\cite{kaplan}. 

 The above results  invalidate any arguments~\cite{dvorn2} in favour of the axial background playing a r\^ole in the generation of instabilities and thus magnification of the magnetic fields in neutron stars. However the KR torsion might play a non-trivial r\^ole in the dynamo equation for the generation of magnetic fields~\cite{dynamo}, and thus affect their strength  independent of the CME. We hope to come back to a discussion of such effects in a future work.

\subsection{The irrelevancy of Chern-Simons terms for the Chiral Magnetic Effect}

In string theory, anomaly cancellation arguments necessitate~\cite{gsw} the modification of the three form representing the field strength of the Kalb-Ramond antisymmetric tensor by gauge (Yang-Mills (Y)) and gravitational (Lorentz (L)) Chern-Simons three forms $\Omega_3$
\begin{equation}\label{cs}
\mathbf{H} = \mathbf{d} \mathbf{B} + \frac{\alpha^\prime}{8\kappa} \Big( \Omega_{\rm 3Y} - \Omega_{\rm 3L}\Big), 
\end{equation}
such that the Bianchi identity (\ref{bianchi}) is now modified to:
\begin{equation}\label{chern}
\mathbf{d} \star \mathbf{H} = \frac{\alpha^\prime}{8\kappa} \, \Big[ {\rm Tr} (\mathbf F \wedge \mathbf F )- {\rm Tr} (\mathbf{R} (\omega) \wedge 
\mathbf{R}(\omega))\Big],
\end{equation}
where now the trace is taken over appropriate gauge and Lorentz indices. This quantity is non-zero if there is no anomaly cancellation. As we discussed previously, by adding appropriate local counterterms to the effective action, one may arrange to add torsion contributions to the right-hand-side of (\ref{chern}), which will appear in a generalised curvature two-form, $\overline{\mathbf{R}}(\overline \omega)$, inside the connection $\overline \omega= \omega + \mathbf{K}$, with $\mathbf{K}$ the contorsion. Equivalently, if one starts (in some effective field theory) from an anomaly equation with torsion, the latter can be removed via the above procedure. 
Thus, unless the gauge and gravitational anomalies are cancelled, the H-torsion charge (\ref{chern}) is not conserved in the presence of Chern-Simons terms. 

In our previous discussion, it was important that we added counterterms to the effective action of string-inspired QED with KR axions, in order to conserve the KR torsion charge in the quantum theory, via the constraint (\ref{delta}) in the path integral; this lead to the absence of torsion from the effective theory (\ref{seffpi}), the associated anomaly equation and to zero contributions of the KR axion to the CME. 
We shall now argue that the latter result is still valid even in the presence of Chern-Simons terms (\ref{chern}).

To this end, it is convenient to use a simplified model, where only a $U(1)$ (electromagnetic gauge group) Chern Simons (CS) term is present. This model was 
considered in \cite{pvqed}, as a string-inspired prototype  of a parity-violating version of quantum electrodynamics with torsion. For our purposes in the current work, we shall restrict ourselves to flat Minkowski space-times, since the introduction of space-time curvature does not affect our conclusions. Let us therefore consider, in the spirit of \cite{pvqed}, the following model~\footnote{This model can be obtained from a string-inspired effective theory of the modified Kalb-Ramond $H$-field with $U(1)$ CS terms, truncated to quadratic order in a derivative expansion, in a curved space-time with 
(non-duynamical) torsion $T_{\mu\nu\rho}$, which appears in both in the Riemann curvature and the fermion gravitational covariant derivative,  and is coupled to the $H_{\mu\nu\rho}$ field via $\int d^4 x \sqrt{-g}\, \kappa^{-1} H_{\mu\nu\rho}\, T^{\mu\nu\rho}$ terms in the action~\cite{pvqed}. This auxiliary torsion field can then be integrated exactly in the path integral, with the result that it obeys the constraint $T_{\mu\nu\rho} = \kappa \, H_{\mu\nu\rho}$. Substituting back to the action, and  taking the Minkowski flat space-time limit, 
leads then to the action (\ref{pvmodel}).}:
\begin{eqnarray}
\label{pvmodel}
{\mathcal S}_{\rm PV} &=& \int d^4x  \Big(-\frac{1}{4} F_{\mu\nu}\, F^{\mu\nu} + \frac{1}{2} \widetilde H_{\mu\nu\rho} \, \widetilde H^{\mu\nu\rho} \Big)
 \nonumber  \\
& +& \int d^{4} \bar{\psi}\Big(\gamma^{\mu}(\imath\, \partial_{\mu}  + q_e \, A_\mu) - m\Big)\psi - \int d^{4}x  {B}_{\mu} \, 
\bar{\psi} \gamma^5 \gamma^\mu \psi 
+ \dots\, 
\end{eqnarray}
where $\widetilde H_{\mu\nu\rho} =\frac{1}{\kappa} \partial_{[\mu}\, B_{\nu\rho]} + \kappa \, A_{[\mu} \, F_{\nu\rho]}$, with $A_\mu$ an electromagnetic $U(1)$ field, $F_{\mu\nu}$ its Maxwell field strength, and in accordance to our previous notation, we work with dimensionless $B_{\mu\nu}$. Here the H-torsion
has dimensions of [mass$^2$], to make contact with the conventions of \cite{pvqed}. The quantity ${B}_\mu$ is the axial torsion pseudovector 
\begin{equation}\label{pseudov}
{B}_\mu = \epsilon_{\mu\nu\rho\sigma}\, \widetilde H^{\nu\rho\sigma}, 
\end{equation}
with $\epsilon_{\mu\nu\rho\sigma}$ the Levi-Civita antisymmetric symbol of flat Minkowski space-time,
and arises from 
identifying the $\widetilde H_{\mu\nu\rho}$ as a (totally antisymmetric) torsion, which thus affects the connection in the  gravitationally covariant derivative of the fermion field. In (\ref{pvmodel}), the ellipsis indicate the repulsive four fermion terms of the form (\ref{seffpi}), characterising every quantum torsion model, when the torsion is integrated out in a path-integral; such terms are suppressed by the Planck mass, and so are ignored for our low-energy analysis. 
The difference of the model (\ref{pvmodel}) from that considered in \cite{pvqed} lies in the fact that here we explicitly consider the charged fermions $\psi$.

The (dimensionless) KR axion $b(x)$ field, is defined in this model from the dual of the $H_{\mu\nu\rho} \equiv \kappa^{-1} \, \partial_{[\mu}\, B_{\nu\rho]} $ term in four space-time dimensions,  $H_{\mu\nu\rho} = \kappa^{-1}\, \epsilon_{\mu\nu\rho\sigma}\, \partial^\sigma b(x)$. In our treatment in previous sections, the latter was identified with the Lagrange multiplier field implementing the path-integral constraint (\ref{delta}) on the conservation of the torsion charge at quantum level~\cite{kaloper}. Such a conservation is spoiled in the specific effective theory (\ref{pvmodel}), due to the Chern-Simons terms. 
Of course, by adding appropriate local counterterms to the effective action (\ref{pvmodel}), one can guarantee the conservation of torsion charge, in which 
case we are led back to the effective theory (\ref{seffpi}), where there are no torsion contributions to the CME. Nonetheless, for the general reader, it will be instructive  to verify explicitly this result in the presence of the Chern-Simons terms, which naively spoil the torsion-charge conservation. 

To this end, one first observes from the action (\ref{pvmodel}) that, to leading order in the inverse Planck-mass suppression factor $\kappa$, 
the axial torsion vector can be approximated by 
\begin{equation}\label{pseudov2}
B_\mu \propto \partial_\mu b + {\mathcal O}(\kappa). 
\end{equation}

As per our previous considerations, we consider a background $ b \to \overline b$, in which the KR axion is linear in (cosmic) time, which yields 
$\overline H_{ijk}= \epsilon_{ijk0}\, \dot{{\overline b}} = {\rm constant}$ as the only non-zero components of $H_{\mu\nu\rho}$. This also implies that only the temporal component ${B}_0$ of the background
pseudovector  (\ref{pseudov2}) is non-vanishing and constant. 

We are now well equipped to proceed with a study of the CME. It is instructive to follow the energy conservation arguments of \cite{kaplan}: 
we consider the rate of change of the total energy of the system (\ref{pvmodel}), that is of the sum of the energy density of the electromagnetic field and of the charged fermions in interaction with the external magnetic fields. We restrict ourselves to flat space-times. To this end, we first note from (\ref{pvmodel}), that the parity-violating interactions of the electromagnetic field with the KR axion are of the form, 
\begin{equation}\label{naivecme}
{\mathcal S}_{\rm PV} \ni   \int d^4x \,  \kappa \, H^{\mu\nu\rho} A_{[\mu}\, F_{\nu\rho]}  = \int d^4x \,  \epsilon^{\mu\nu\rho\sigma} \, \partial_\sigma b \, A_{[\mu}\, F_{\nu\rho]} .
\end{equation}
One might plausibly reason that, in the background $\dot{\overline{b}} \equiv B_0 = {\rm constant}$, the terms (\ref{naivecme}) would yield a CME-like effect,
with the current being proportional to $B_0$. In the presence of an external magnetic field $\vec{\mathcal B}$ in that case, one would obtain the parity-violating action terms~\footnote{The overall minus sign is due to the fact that the three-dimensional Levi-Civita symbol $\epsilon^{ijk}$ is defined as: $\epsilon^{ijk}= \epsilon^{0ijk} = - \epsilon^{ijk0}$.}: ${\mathcal S}_{\rm PV} \ni -\int d^4 x \, B_0 \, \epsilon^{ijk} A_i F_{jk} = - 2\int d^4 x B_0 \,\vec A \cdot \vec{\mathcal B}$. However, this naive expectation is not correct, as we now explain.

 The parity-violating interaction terms (\ref{naivecme}) contribute to the energy density of the electromagnetic field, which, to leading order in $\kappa$ and for the constant $\dot{\overline{b}}$ KR axion background, assumes the form:
\begin{equation}\label{emed}
{\mathcal E}_{em} = \frac{1}{2} \big(\vec E^2 + \vec{\mathcal B}^2 \big)  + 2 \,  \,  \dot{\overline b}\, \vec A \cdot \vec{\mathcal B}.
\end{equation}
The first term in parenthesis on the right-hand side is the standard Maxwell electrodynamics term. 

In the context of the conventional CME~\cite{cme}, we consider homogeneous electric $\vec E$ and magnetic $\vec{\mathcal B}$ fields,  and in fact a constant intensity of the magnetic field $\vec B$. Under these conditions, the rate of change of the energy density $ \mathcal{E}_{em}$ (\ref{emed}) is:
\begin{equation}\label{rce}
\frac{d}{dt} \mathcal{E}_{em} = \vec E \cdot \frac{d\vec E}{dt} + 2\, \, B_0  \,  \vec E \cdot \vec{\mathcal B}.
\end{equation}
It is important that at this stage we consider temporal dependence of the electric field. Eventually we shall consider 
$\vec E \to 0$, parallel to the externally applied magnetic field, which are the conditions in which the CME is conventionally discussed~\cite{cme}. 

On the other hand, as we mentioned above~\cite{kaplan,dvorn}, the energy of charged fermions ${\mathcal E}_{\rm fermion}$ interacting with a constant external magnetic field $\vec B$ is determined 
by means of the Landau levels, of which only the lowest $n=0$, contributes to the CME current. In the presence of a chemical potential $\mu_5$, the rate of the corresponding energy density
is given by 
\begin{equation}\label{energye}
 \frac{d}{dt} {\mathcal E}_{\rm fermion} = \frac{2\alpha}{\pi}\mu_5\, \vec E \cdot \vec{\mathcal B},
\end{equation}
and has its origin in the chiral anomaly~\cite{cme,kaplan}. Notice that the axial torsion background ${B}_0$ (\ref{pseudov2}), 
does not contribute to this energy, as already mentioned~\cite{kaplan,dvorn}. 

Energy conservation requires~\cite{kaplan}:
\begin{equation}\label{enercons}
\frac{d}{dt}{\mathcal E}_{em} + \frac{d}{dt } {\mathcal E}_{\rm fermion} = 0.
\end{equation}
From the equations of motion of the electromagnetic potential $A_\mu$ obtained from the action (\ref{pvmodel}), in our case, the only non-trivial one reads:
\begin{equation}\label{maxwell}
\frac{d}{dt} \vec E + 2 B_0 \, \vec{\mathcal B} = - \vec j_q
\end{equation}
where $\vec j_q = q_e \bar \psi \vec \gamma \psi $, is the charged fermion current (with the notation $\vec \gamma $ referring collectively to the spatial Dirac matrices). 

On account of (\ref{maxwell}), (\ref{enercons}), (\ref{energye})  and (\ref{rce}),  we obtain 
\begin{equation}\label{final}
\Big(\vec j_q - \frac{2\alpha}{\pi} \mu_5 \vec{\mathcal B} \Big) \cdot \vec E =0,
\end{equation}
which for an infinitesimal electric field $\vec E \to 0$ parallel to the constant magnetic field, yields the standard CME effect (\ref{chiral3}). Notice here that had one naively set $\vec E =0$ in the Maxwell equation (\ref{maxwell}), one would have erroneously derived a CME-like current $\vec j_q =-2 B_0 \, \vec{\mathcal B}$ induced by the torsion background. This would violate energy conservation. What is actually happening here, is that the quantum fluctuations of the electric field produce a $d\vec E/dt \ne 0$, and yield contributions to both the kinetic and the potential energy of the electrons, that cancel any effects of $B_0$ on the induced current, which thus depends only on the chemical potential (\ref{final}), (\ref{chiral3}). 
This concludes the proof of our statement that the presence of the Chern-Simons terms (\ref{chern}) does not alter the conclusion that the torsion cannot 
contribute to the chiral magnetic effect. More formally the torsion terms can be removed  from the chiral anomaly by addition of appropriate local counterterms to the effective action~\cite{hull,mavindex}, and thus they cannot contribute to phenomena associated with the anomaly, such as the CME.

\section{KR axions and anomalous  generation of Majorana mass for the Right-handed neutrinos \label{sec:mass}}

In our discussion so far on leptogenesis, we have assumed a bare mass $m_N$ for RHN (\ref{smelag}), without providing any discussion on its microscopic origin. In standard scenarios, such a mass scale is provided by the seesaw mechanism~\cite{seesaw}. We now review an approach to  mass generation of RHN~\cite{pilaftsis} which constitutes a novel mechanism for neutrino mass generation  beyond the conventional seesaw framework~\cite{seesaw}. The mechanism involves  quantum fluctuations of  a KR axion and a quantum anomaly.

An important aspect of the coupling  of  the KR axion quantum field $b(x)$) to
the  fermionic   matter in the action (\ref{seffpi})  is  its   \emph{shift  symmetry},
characteristic of an axion field. By shifting the field $b(x)$
by  a constant:  $b(x)  \to b(x)  +  c$, the  action only
changes by  total derivative terms, such  as $c\, R^{\mu\nu\rho\sigma}
\widetilde{R}_{\mu\nu\rho\sigma}$               
and $c\, F^{\mu\nu}\widetilde{F}_{\mu\nu}$.  These terms are irrelevant for the
equations  of motion and  the induced  quantum dynamics,  provided the
fields fall off sufficiently fast  to zero at space-time infinity. 
The scenario for  the anomalous  Majorana mass generation  through torsion proposed in \cite{pilaftsis}, which we shall review briefly below,  
consists of augmenting the  effective action (\ref{seffpi}) by terms that
break such a shift symmetry. 

To this end, we  first couple the KR axion $b(x)$ to
another pseudoscalar  axion field $a(x)$.   In string-inspired models,
such pseudoscalar   axion~$a(x)$  may  be  provided   by  the  string
moduli~\cite{arvanitaki}. The  proposed   coupling  occurs
through  a mixing  in  the kinetic  terms  of the  two  fields. To  be
specific, we consider the action:
\begin{eqnarray} 
  \label{bacoupl}
    \mathcal{S} \!&=&\! \int d^4 x \sqrt{-g} \, \Big[\frac{1}{2}
      (\partial_\mu b)^2 + \frac{b(x)}{192 \pi^2 f_b}
      {R}^{\mu\nu\rho\sigma} \widetilde{R}_{\mu\nu\rho\sigma} 
      + \frac{1}{2f_b^2} J^5_\mu {J^5}^\mu + \gamma
      (\partial_\mu b )\, (\partial^\mu a ) + \frac{1}{2}
      (\partial_\mu a)^2\nonumber\\ 
&&- i y_a \, a(x) \, \Big( \overline{\psi}_R^{\ C} \psi_R - \overline{\psi}_R
\psi_R^{\ C}\Big) \Big] + \dots \;, \qquad f_b = \Big(\frac{3\, \kappa^2}{8}\Big)^{-1/2} = \frac{M_P}{\sqrt{3\, \pi}}
\end{eqnarray}
where the $\dots$ indicate terms in the low-energy string effective action, including SM ones, that are not of direct relevance to our purposes in the present article. The anomaly equation (\ref{anomalygrav2}) has been used to yield the second term on the right hand side of (\ref{bacoupl}), by partial integration of the corresponding $\partial b(x)-J^5$ term of the action (\ref{seffpi}). 
Here, we have ignored gauge  fields, which are
not of interest to us, given that there is no direct coupling of RHN to them.
Moreover, for our purposes, the form of the axion $a(x)$ potential, including details of its mass $M_a$,  are not relevant~\cite{pilaftsis}. Above, $\psi_R^{\ C} $ is the charge-conjugate right-handed
fermion $\psi_R$, with the corresponding four-component Majorana spinor being defined as $\psi =
\psi_R  +  (\psi_R)^C$,  and  $\gamma$  is  a  real  parameter  to  be
constrained later on.  The Yukawa
coupling $y_a$ of  the axion moduli field $a$  to right-handed sterile
neutrino  matter $\psi_R$  may  be due  to  non perturbative  effects (e.g. string instantons), 
which are known (along with the axion potential itself) to\emph{break} the shift symmetry: $a \to a + c$.

It is convenient to diagonalize  the axion kinetic terms in (\ref{bacoupl}) by redefining
the KR axion field as follows:
$b(x) \rightarrow {b^\prime}(x) \equiv b(x) + \gamma a(x)$. It can then be easily seen~\cite{pilaftsis} that 
the $b^\prime  $ field decouples and can be
integrated out  in the  path integral, leaving  behind an  axion field
$a(x)$  coupled   both  to  matter   fermions  and  to   the  operator
$R^{\mu\nu\rho\sigma}   {\widetilde   R}_{\mu\nu\rho\sigma}$.  As discussed in \cite{pilaftsis}, though, 
the approach is only valid for
$
|\gamma|\ <\ 1\; , $
otherwise the  axion field would  appear as a  ghost, \emph{i.e.}~with
the  wrong  sign  of  its  kinetic  terms,  which  would  indicate  an
instability  of  the  model.  This  is the  only  restriction  of  the
parameter $\gamma$. In this case we may redefine the axion field so as to appear with a
canonical normalised kinetic term, implying the effective action:
\begin{eqnarray} 
  \label{bacoupl3} 
\mathcal{S}_a &= \int d^4 x
    \sqrt{-g} \, \Big[\frac{1}{2} (\partial_\mu a )^2 - \frac{\gamma
        a(x)}{192 \pi^2 f_b \sqrt{1 - \gamma^2}}
      {R}^{\mu\nu\rho\sigma} \widetilde{R}_{\mu\nu\rho\sigma}  \nonumber \\ &- i\, \frac{y_a}{\sqrt{1 - \gamma^2}} \, 
a(x)\, \Big( \overline{\psi}_R^{\ C} \psi_R - \overline{\psi}_R
\psi_R^{\ C}\Big) + \frac{1}{2f_b^2} J^5_\mu {J^5}^\mu
      \Big]\; .
\end{eqnarray}
Evidently, the  action $\mathcal{S}_a$ in~(\ref{bacoupl3}) corresponds
to a  canonically normalised axion field $a(x)$,  coupled \emph{both }
to the curvature of space-time, \emph{\`a la} torsion, with a modified
coupling $\gamma/(192 \pi^2  f_b \sqrt{1-\gamma^2})$, and to fermionic
matter  with  chirality-changing  Yukawa-like  couplings of  the  form
$y_a/\sqrt{1 - \gamma^2}$.
\begin{figure}[t]
 \centering
  \includegraphics[clip,width=0.40\textwidth,height=0.15\textheight]{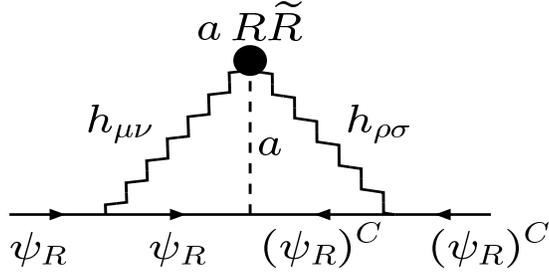} 
\caption{\it Two-loop Feynman graph giving rise to anomalous generation of Majorana
  mass for the right-handed neutrinos $\psi_R$~\cite{pilaftsis}.  The black circle denotes the operator $a(x)\,
  R_{\mu\nu\lambda\rho}\widetilde{R}^{\mu\nu\lambda\rho}$ induced by anomalies (at one-loop). The fields $h_{\mu\nu}$ (wavy lines) denote graviton fluctuations. Straight lines with arrows denote right handed neutrino fields and their conjugates.}\label{fig:feyn}
\end{figure}

The mechanism for  the anomalous Majorana mass generation  is shown in
Fig.~\ref{fig:feyn}. Only graviton fluctuations and axion $a(x)$ fields couple to the RHN at leading orders. We may estimate  the  two-loop Majorana
mass for the RHN, $M^N_R$,  in  quantum gravity with an effective  UV energy cut-off
$\Lambda$, by  adopting  the   effective   field-theory  framework   of
\cite{Donoghue:1994dn}:
\begin{equation}
  \label{MR}
M^N_R \sim 
\frac{\sqrt{3}\, y_a\, \gamma\,  \kappa^5 \Lambda^6}{49152\sqrt{8}\,
\pi^4 (1 - \gamma^2 )}\; .  
\end{equation} 
In a UV
complete theory  such as  strings, the cutoff $\Lambda$  and the Planck mass scale $M_P$  are related.

If the cut-off $\Lambda$ is of the same order as the reduced Planck mass of the four-dimensional theory, that is $\kappa \, \Lambda \sim 1$, then from (\ref{MR}), we observe that for $y_a \, \gamma ={\mathcal O}(10^{-6})$ one obtains $M^N_R \sim 10^5$~GeV, of the order of the RHN in our leptogenesis scenario. On then other hand, for much lower $\Lambda$ of order of the GUT scale,  $\Lambda \sim 10^{16}$~GeV, one obtains RHN masses $M^N_R ={\mathcal O}(16)$~keV, that is in the warm dark matter regime of the $\nu$MSM~\cite{Shaposhnikov:2006xi}, consistent with current stringent constraints~\cite{perez,Yunis}. The mass hierarchy among the RHN is arranged by appropriate choices of the Yukawa couplings $y_a$, $a=1,2,3$, in that case.

We now remark that in string theory there are several axion-like (pseudoscalar) fields $a_i(x)$, $i=1,2, \dots n$, originating from flux fields that exist in the spectrum~\cite{arvanitaki}, in addition to the aforementioned $B_{\mu\nu}$ Kalb-Ramond field. One can then assume~\cite{pilaftsis} the existence of Yukawa couplings with right-handed neutrinos, provided some non-perturbative instanton effects are responsible for a breaking of the shift symmetry. These string-theory axion fields could
mix  with each  other.  Such  a  mixing can  give rise  to reduced  UV
sensitivity of  the two-loop  graph shown in  Fig.~\ref{fig:feyn}.  To
make this point explicit, let  us consider a scenario with $n$ axion fields, $a_{1,2,\dots,n}$, of which 
only $a_1$ has a  kinetic mixing term $\gamma$ with the
KR  axion  $b$  and  only   $a_n$  has  a  Yukawa  coupling  $y_a$  to
right-handed neutrinos  $\psi_R$.  The other  axions $a_{2,3,\dots,n}$
have a next-to-neighbour mixing pattern.  In such a model, the kinetic
terms of the effective action are given by {\small
\begin{eqnarray}
  \label{Skin}
\mathcal{S}^{\rm  kin}_a \, = \,  \int  d^4x\, \sqrt{-g}\,  \bigg[
  \frac{1}{2}\,\sum\limits_{i=1}^n \Big( (\partial_\mu   a_i)^2 - M^2_i\Big)  
+ \gamma  (\partial_\mu
  b)(\partial^\mu a_1)  - \frac{1}{2}\,\sum\limits_{i=1}^{n-1} \delta M^2_{i,i+1}\,
a_i a_{i+1}\ \bigg]\; ,
\end{eqnarray}
\hspace{-1.5mm}}
where the mixing mass terms $\delta M^2_{i,i+1}$ are constrained to be
$\delta M^2_{i,i+1} < M_i M_{i+1}$,  so as to obtain a stable positive
mass   spectrum   for  all   axions.    As   a   consequence  of   the
next-to-nearest-neighbour mixing, the UV behaviour of the off-shell transition
$a_1 \to a_n$, described by the propagator matrix element $\Delta_{a_1
  a_n} (p)$,  changes drastically, i.e.~$\Delta_{a_1  a_n} (p) \propto
1/(p^2)^n \sim  1/E^{2n}$.  Assuming,  for simplicity, that  all axion
masses  and  mixings  are  equal,  i.e.~$M^2_i =  M^2_a$  and  $\delta
M^2_{i,i+1} =  \delta M^2_a$, the anomalously  generated Majorana mass
may be estimated to be
\begin{equation}
  \label{MRmix1}
M^N_R \sim 
\frac{\sqrt{3}\, y_a\, \gamma\,  \kappa^5 \Lambda^{6-2n} 
(\delta M^2_a)^n}{49152\sqrt{8}\, 
\pi^4 (1 - \gamma^2 )}\; ,
\end{equation}
for $n \leq 3$, and 
\begin{equation}
  \label{MRmix2}
M^N_R \sim 
\frac{\sqrt{3}\, y_a\, \gamma\,  \kappa^5 (\delta M^{2}_a)^3}{49152\sqrt{8}\, 
\pi^4 (1 - \gamma^2 )}\; \frac{(\delta
  M^{2}_a)^{n-3}}{(M^2_a)^{n-3}}\; ,
\end{equation}
for  $n >  3$.  It  is then  not difficult  to see  that  three axions
$a_{1,2,3}$ with next-to-neighbour mixing  as discussed above would be
sufficient  to obtain a  UV finite  (cut-off-$\Lambda$-independent) result for  $M^N_R$ at  the two-loop
level. Of  course, beyond the two  loops, $M_R$ will  depend on higher
powers of the energy  cut-off $\Lambda$, i.e.~$\Lambda^{n> 6}$, but if
$\kappa\Lambda \ll  1$, these higher-order effects are  expected to be
subdominant.

In the above $n$-axion-mixing  scenarios, we note that the anomalously
generated  Majorana mass  term  will only  depend  on the  mass-mixing
parameters $\delta M_a^2$ of the  axion fields and not on their masses
themselves, as long as $n \le 3$.  Instead, for axion-mixing scenarios
with $n > 3$, the induced Majorana neutrino masses are proportional to
the factor $(\delta M^2_a/M^2_a)^n$, which gives rise to an additional
suppression for heavy axions with masses $M_a \gg \delta M_a$.

\section{Conclusions  \label{sec:concl}} 

In this work we have reviewed our previous studies of leptogenesis induced by  a (rather generic) LV and CPTV time-like axial background  $B_0$, 
which is provided by the field strength of the antisymmetric tensor KR field appearing in the massless spectrum of microscopic, ultraviolet complete, string-inspired models. We discussed briefly scenarios where the background $B_0$ 
is either constant, for a given epoch of the universe, or 
varying slowly with the temperature of the early universe. The phenomenology of leptogenesis, associated with a lepton asymmetry generated by the (asymmetric) decays of heavy right-handed Majorana neutrinos (RHN)  into SM leptons and antileptons, in the presence of temperature-dependent axial backgrounds, remains largely unchanged from the constant background case, and is consistent with the stringent current-era epoch constraints on LV and CPTV. Fine tuning however is required so as to ensure the suppression of the vacuum energy density. This can be provided by bulk anti-de-Sitter contributions to the vacuum energy density of our Universe in brane models, where our world is viewed as a three brane propagating in a higher dimensional bulk space time. 

As a byproduct of our analysis, we also argued that the background $B_0$ does not contribute to the so-called Chiral Magnetic Effect (CME), that is the induction of an electrical current proportional to external magnetic fields, which characterises physical systems in the presence of chiral chemical potentials, $\mu_5$ (i.e. where there is a non-zero difference, $\mu_L-\mu_R \ne 0$, between the chemical potentials of left- and right-handed chiral fermions). Since, during the leptogenesis era one may have encountered primordial magnetic fields, and given the apparent r\^ole of the background $B_0$ in the effective Lagrangian as a dynamical contribution to $\mu_5$, it is natural to ask whether $B_0$ contributes to CME. As we argued above, this is not the case. This result can be understood either from the point of view of a generic axial background, which is known for energetic reasons not to contribute to CME, or, of our string-inspired model, from the fact that the background $B_0$ is associated with the (totally antisymmetric) ``H-torsion'' induced by the KR field. The CME is associated with the chiral anomaly. Also it is a well known that in string effective theories the H-torsion contributions to the anomaly can be removed by a choice of the renormalisation scheme. So the null contribution of the H-torsion to the CME should be expected on these grounds. 

Finally, the r\^ole of quantum fluctuations of the KR axion in generating an anomalous Majorana mass for the RHN themselves, was also described. For this latter scenario, a kinetic mixing between the KR axion with ordinary axion fields $a(x)$ (QCD or string inspired) is assumed, together with shift-symmetry breaking chirality-changing Yukawa interactions of the axions $a(x)$ with the Majorana neutrinos (which might be the result of non-perturbative effects (instantons) in microscopic string models). This mechanism is novel and goes beyond the conventional seesaw. It is interesting to remark that, within our string framework, it is known that there can be several axion fields, which allows an ultraviolet-cutoff independent RHN Majorana mass for the special case of three axion fields and three RHNs. 

It would be interesting to pursue further the detailed phenomenology and cosmology of such models, in particular to examine the effects of temperature in the induced RHN mass, and thus embed fully this mechanism within our CPTV leptogenesis scenario. This will be investigated in a future work.

\section*{Acknowledgments}

This research was funded in part by STFC (UK) research grant ST/P000258/1. The article is based on an invited plenary talk given by N.E.M. at the International Conference of New Frontiers in Physics 2018 (Kolymbari, Crete (Greece)). 
The authors wish to thank the organisers of the conference for organising such a high-level and stimulating event. N.E.M. also acknowledges a scientific associateship (``\emph{Doctor Vinculado}'') at IFIC-CSIC-Valencia University\ (Spain).

\end{document}